\documentclass[twocolumn,colorlinks,showpacs,preprintnumbers,,superscriptaddress]{revtex4}
\usepackage{graphicx}
\usepackage{epsf}
\usepackage{color}
\usepackage{dcolumn}
\usepackage{bm}
\usepackage{verbatim}
\usepackage{amssymb}
\begin{document}

 
\title{Field driven magnetostructural transitions in GeCo$_2$O$_4$}
\author{X. Fabr\`eges}
\affiliation{Laboratoire L\'eon Brillouin, CEA, CNRS, Universit\'e Paris-Saclay, CEA Saclay, F-91191 Gif-sur-Yvette Cedex France}
\affiliation{Laboratoire National des Champs Magn\'etiques Intenses, CNRS-INSA-UJF-UPS, 31400 Toulouse, France}
\author{E. Ressouche}
\affiliation{CEA/Grenoble, INAC/SPSMS-MDN, 17 rue des Martyrs, 38054 Grenoble Cedex 9, France}
\author{F. Duc}
\affiliation{Laboratoire National des Champs Magn\'etiques Intenses, CNRS-INSA-UJF-UPS, 31400 Toulouse, France}
\author{S. de Brion}
\affiliation{CNRS, Institut N\'eel, 38042 Grenoble, France}
\affiliation{Univ. Grenoble Alpes, Institut N\'eel, 38042 Grenoble, France}
\author{M. Amara}
\affiliation{CNRS, Institut N\'eel, 38042 Grenoble, France}
\affiliation{Univ. Grenoble Alpes, Institut N\'eel, 38042 Grenoble, France}
\author{C. Detlefs}
\affiliation{European Synchrotron Radiation Facility, 71 Avenue des Martyrs, 38000 Grenoble, France}
\author{L. Paolasini}
\affiliation{European Synchrotron Radiation Facility, Grenoble, France}
\author{E. Suard}
\affiliation{Institut Laue Langevin, BP 156, 38042 Grenoble Cedex 9, France}
\author{L.-P. Regnault}
\affiliation{CEA/Grenoble, INAC/SPSMS-MDN, 17 rue des Martyrs, 38054 Grenoble Cedex 9, France}
\author{B. Canals}
\affiliation{CNRS, Institut N\'eel, 38042 Grenoble, France}
\author{P. Strobel}
\affiliation{CNRS, Institut N\'eel, 38042 Grenoble, France}
\affiliation{Univ. Grenoble Alpes, Institut N\'eel, 38042 Grenoble, France}
\author{V. Simonet}
\email[Corresponding author: ]{virginie.simonet@neel.cnrs.fr}
\affiliation{CNRS, Institut N\'eel, 38042 Grenoble, France}
\affiliation{Univ. Grenoble Alpes, Institut N\'eel, 38042 Grenoble, France}

\date{\today}

\begin{abstract}
In the spinel compound GeCo$_2$O$_4$, the Co$^{2+}$ pyrochlore sublattice presents remarkable magnetic field-induced behaviors that we unveil through neutron and X-ray single-crystal diffraction. The N\'eel ordered magnetic phase is entered through a structural lowering of the cubic symmetry. In this phase, when a magnetic field is applied along a 2-fold cubic direction, a spin-flop transition of one fourth of the magnetic moments releases the magnetic frustration and triggers magnetostructural effects. At high field, these ultimately lead to an unusual spin reorientation associated with structural changes.  
\end{abstract}

\pacs{75.25.-j,75.50.Ee,75.30.Gw,75.80.+q}

\maketitle

\section{Introduction}

Spinel compounds with the generic formula $AB_2\mathrm{O}_4$ crystallize in the cubic space group $Fd\bar{3}m$. The $B$ sites can accomodate a magnetic ion, in this case $\mathrm{Co}^{2+}$. They form a pyrochlore lattice, a network of corner-sharing tetrahedra, that is the archetype of geometrical frustration. In addition, they combine magnetic and lattice degrees of freedom, which confer them a magnetostructural flexibility in zero and finite magnetic field. The interplay between magnetic frustration and magnetoelastic coupling has been intensively studied in spinels with, for instance, either V$^{3+}$ or Cr$^{3+}$ on the $B$ site and various ions on the $A$ site (Hg, Mg, Cd and Zn) \cite{Lee2000,Ueda2005,Ueda2006,Matsuda2007,Matsuda2010,Mun2014}. In these systems, at the N\'eel temperature ($T_{\rm N}$), a transition to an antiferromagnetic (AFM) ordering is accompanied by a cubic to tetragonal or orthorhombic structural distortion. It is interpreted as a 3-dimensional spin-Peierls transition acting to reduce the frustration \cite{Lee2000,Ueda2006}. In the Cr compounds, when applying a magnetic field, a magnetization plateau at half of the saturation magnetization is stabilized by the spin-lattice coupling on a wide range of fields \cite{Ueda2005, Ueda2006,Penc2004,Bergman2006}. Moreover, the beginning of the plateau coincides with a structural distortion. It corresponds to a recovery of the cubic structure as a consequence of the release of the frustration by the magnetic field \cite{Matsuda2007,Matsuda2010}. 

\begin{figure*}
\begin{center}
\includegraphics[width=17cm]{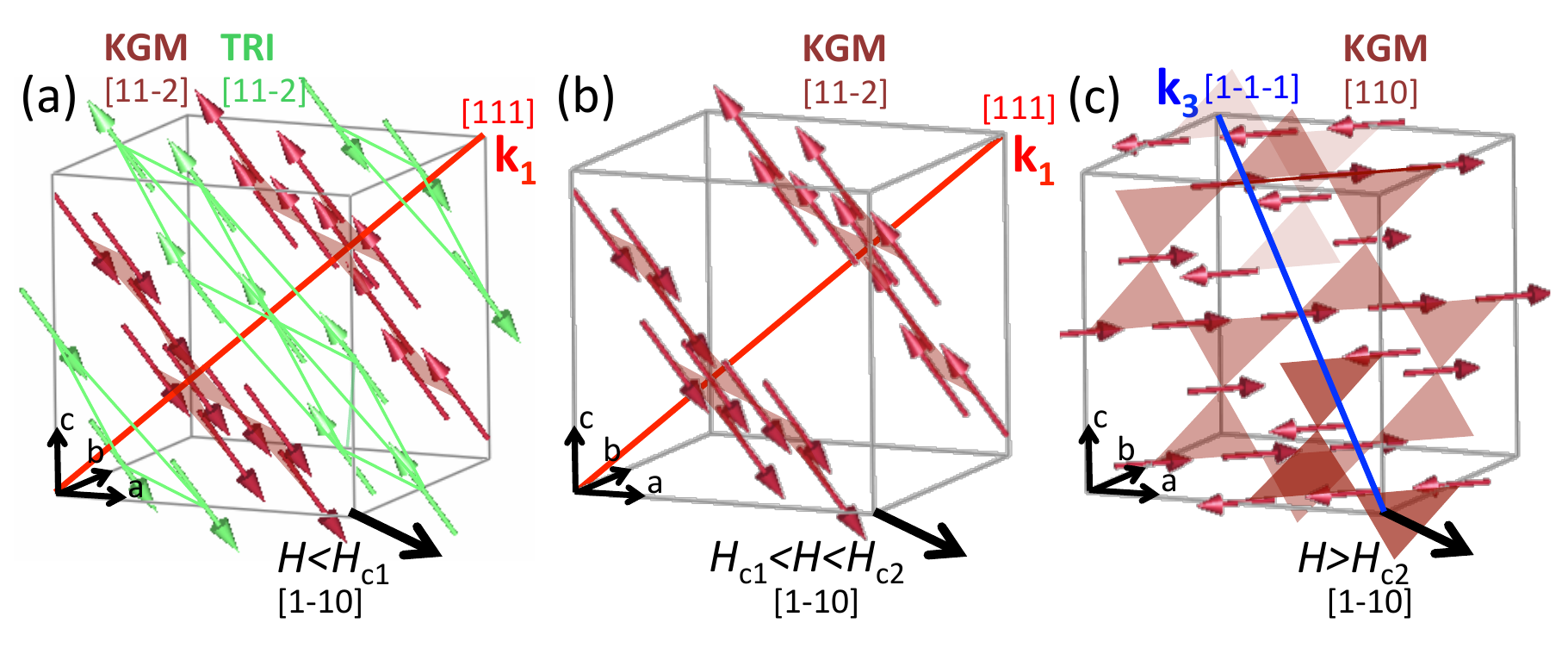}
\end{center}
\caption{AFM ordered components of the KGM (red) and TRI (green) moments determined from low temperature single-crystal neutron diffraction refinements in zero-field and under a magnetic field applied along the [1-10] direction (black arrow): at $\mu_0H$=0 T for the $\bf{k_1}$ domain (one of the 3 equipopulated $S$-domains represented) (a), at $\mu_0H$=5.5 T for the $\bf{k_1}$ domain (b) and at $\mu_0H$=12 T for the $\bf{k_3}$ domain after the high-field spin-reorientation (c). Note that the ferromagnetic component that develops under magnetic field at the expense of the antiferromagnetic component is not shown. The AFM component of the TRI magnetic moments is only present in (a) since they get polarized by the magnetic field above $\it{H_{\rm c1}}$. From (b) to (c), the KGM planes, in which lie the AFM component of the magnetic moments, are reoriented from (111) to (1-1-1), consistent with the change of propagation vector.  }
\label{Refin}
\end{figure*}

Ge spinels, with Ni$^{2+}$ or Co$^{2+}$ on the pyrochlore lattice, have comparable rich phase diagrams but have been less studied because the frustration effects are not as straightforward, in particular not solely driven by the first neighbor interactions. They were both shown, by powder neutron diffraction, to order in a magnetic structure characterized by a propagation vector $\bf k$=(1/2, 1/2, 1/2) and stabilized through competing interactions beyond the third neighbors \cite{Diaz2006,Matsuda2011}. The magnetic arrangement was described assuming a single propagation vector (single-$k$ structure) as follows: alternating kagome (KGM) and triangular (TRI) ferromagnetic (FM) planes, perpendicular to the $<$111$>$ directions associated with the propagation vector, and antiferromagnetically coupled to the nearest planes of the same kind (see figure \ref{Refin}a). In GeNi$_2$O$_4$, which retains the cubic structure in the magnetic phase, both KGM and TRI sites are magnetically decoupled, i.e. the molecular field created by the magnetization of one type of site on the other one is zero. It leads to an independent ordering of the two kinds of planes with distinct transition temperatures \cite{Bertaut,Crawford2003,Matsuda2008}. This is at variance with GeCo$_2$O$_4$ in which a unique magnetic transition is observed at $T_{\rm N}$=23.5 K. This is ascribed to a structural distortion that couples the two sites and allows to reduce the frustration in the absence of any external magnetic field. This distortion was proposed to be mainly cubic-to-tetragonal from powder neutron and X-ray diffraction \cite{Hoshi2007,Barton2014}.

Besides this zero-field magnetostructural behavior, a complex $H$-$T$ phase diagram has been observed in GeCo$_2$O$_4$ by magnetization, ultrasound and electron spin resonance measurements, evidencing several field-induced anomalies \cite{Plumier1967,Diaz2004,Diaz2005,Diaz2006,Hoshi2007,Sasame2007,Watanabe2011,Yamasaki2013,Tomiyasu2011b}. Their microscopic origin was investigated by powder neutron diffraction under magnetic field up to 10 T \cite{Matsuda2011}. A transition, observed at $\approx$ 4 T, was attributed to an antiferromagnetic-to-ferromagnetic  spin rearrangement between the triangular planes while retaining the antiferromagnetic arrangement between the kagome planes. The final ferromagnetic order of the kagome planes was proposed to occur at the second transition around $\mu_0H$=10 T.

We hereafter present the results of single-crystal neutron and synchrotron X-ray diffraction under zero and finite magnetic fields applied along a 2-fold axis of the high temperature cubic structure. We show in particular that the high field transition, for this orientation of the field, is much more subtle than a plain antiferromagnetic-to-ferromagnetic one. It implies a change of the magnetic anisotropy, a switch of the magnetostructural domains, and the stabilization of a new canted magnetic structure. We discuss the interplay between magnetostructural effects and frustration in triggering these field-induced unconventional behaviors. 

\section{Experimental details}

Single-crystals of GeCo$_2$O$_4$ with an octahedral shape of approximately 2 mm size were synthesized by chemical vapor transport in 0.1 atm of HCl gas. A heat treatment was performed with a heating plateau at 950$^{\circ}$C followed by a slow cooling at 1$^{\circ}$/mn. 

Magnetization measurements on the oriented single-crystal were performed up to 10 T in the temperature range 2K -300 K using an extraction magnetometer. The magnetic susceptibility data are in agreement with those published in the literature with a N\'eel temperature of 23.5 K (see Supplemental Material).

Single crystal neutron diffraction was carried out at the Institut Laue Langevin (ILL). Single-crystal diffraction in zero magnetic field was obtained on the CEA-CRG D15 diffractometer operated in the 4-circle mode, with the sample mounted in a displex refrigerator, and with an incident wavelength of 1.173 \AA. Measurements under magnetic field up to 12 T was performed on the CEA-CRG D23 two-axis diffractometer with a lifting arm detector and an incident wavelength of 1.280 \AA. The single-crystal was mounted in the cryomagnet with the [1-10] axis set vertical, parallel to the applied field. An additional diffraction experiment using polarized neutrons with spherical polarization analysis was performed on the single-crystal installed inside the CRYOPAD device on the CEA-CRG IN22 spectrometer at the ILL using a wavelength of 2.36 \AA. This set-up allows measuring the three orthogonal components of the polarization vector of the neutron beam after scattering by the sample whatever the direction of the incident neutron beam polarization. It gives for a reflection ($Q_h$, $Q_k$, $Q_l$) a polarization matrix $\overline{P_{ij}}$, with $i$,$j$=X,Y,Z (X $\parallel$ scattering vector, Z $\perp$ the scattering plane) \cite{Tasset1989,Brown1993}. 

\begin{figure}
\begin{center}
\includegraphics[width=8.5cm]{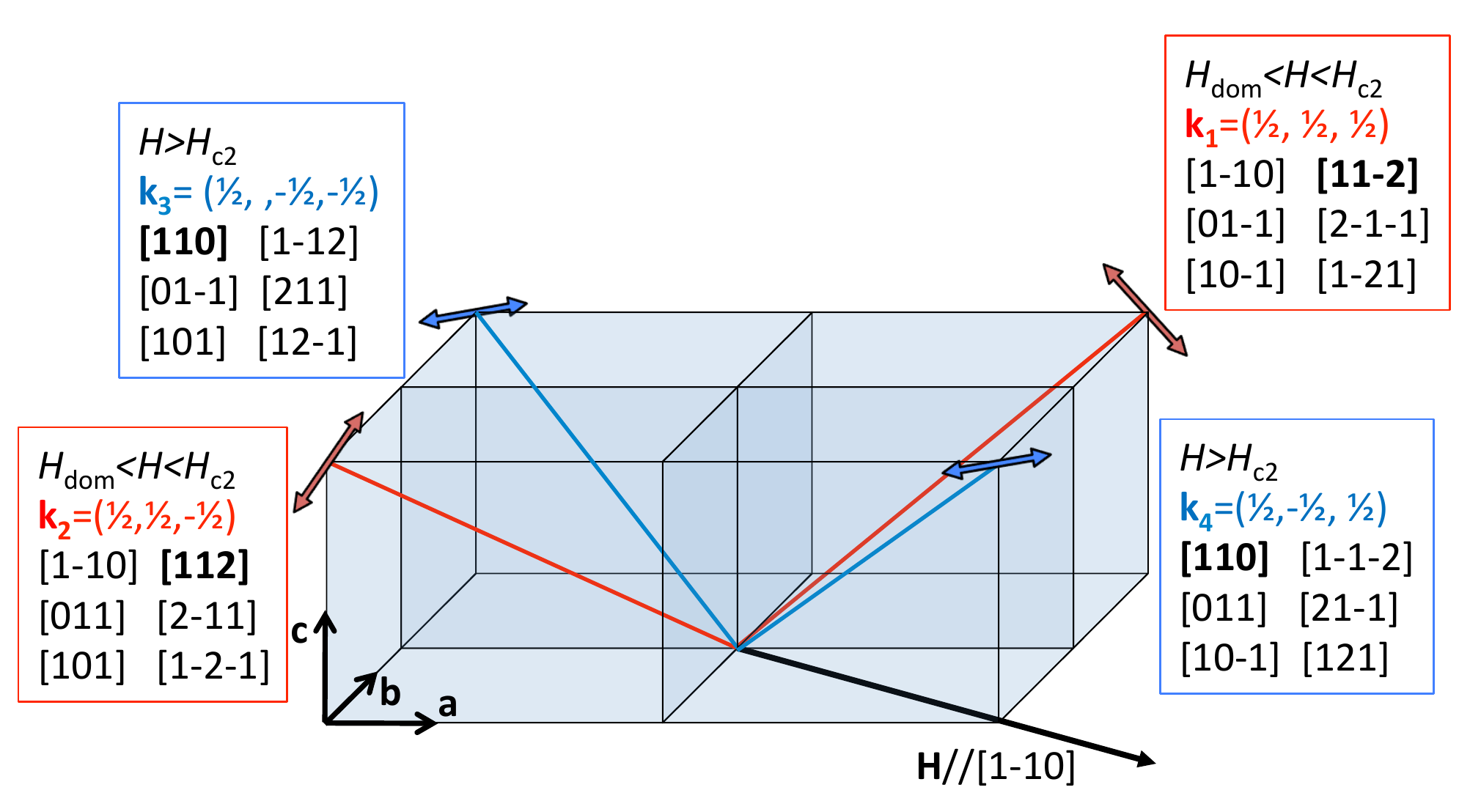}
\end{center}
\caption{Magnetic domains in cubic symmetry: the direction of the propagation vectors is shown by red lines for the pair ($\bf{k_1}$, $\bf{k_2}$) selected between $\it{H_{\rm dom}}$ and $\it{H_{\rm c2}}$, and by blue lines for the other pair ($\bf{k_3}$, $\bf{k_4}$) selected above $\it{H_{\rm c2}}$. For each $k$-domain, the 3 $S$-domains are listed in the boxes for two possible orientation of the magnetic moments, either along the $<$110$>$ directions (left) or along the $<$112$>$ directions (right). The domains in bold and the double arrows give the orientation of the AFM component on the KGM sites for $\it{H_{\rm dom}}$ $<$ $\it{H}$ $<$ $\it{H_{\rm c2}}$ and for $\it{H}$ $>$ $\it{H_{\rm c2}}$.}
\label{MagDom}
\end{figure}

High magnetic field X-ray single-crystal diffraction experiments were performed at the ID06 beamline of the European Synchrotron Radiation Facility using a monochromatic beam selected by a Si(111) double monochromator. Magnetic field was provided by a 10 T split-coil superconducting magnet \cite{Paolasini2007}. A 2x2x1\,mm single-crystal was aligned with the [110] and [001] axes within the horizontal plane giving access to the (\emph{h,h,l}) Bragg reflections. Magnetic field was applied along the [1-10] direction. The measurements were performed with $E=32$\,keV ($\lambda=0.3876$\,\AA)  and the diffracted intensities were collected with a photodiode. 

\section{Results}


\subsection{Magnetic structure in zero and finite magnetic fields}

To analyze the single-crystal neutron diffraction results of GeCo$_2$O$_4$, we first assume a cubic structure and a single-$k$ magnetic arrangement. Both hypotheses will be discussed later on. We take into account the presence of several magnetic domains, that appear at the phase transition (see figure \ref{MagDom}): we consider four $k$-domains that correspond to the symmetry-equivalent directions of propagation of the AFM order given by the propagation vectors $\bf{k_1}$=(1/2, 1/2, 1/2), $\bf{k_2}$=(1/2, 1/2, -1/2), $\bf{k_3}$=(1/2, -1/2, -1/2), $\bf{k_4}$=(1/2, -1/2, 1/2). Moreover, if the magnetic moments lie in the $\{$111$\}$ planes, there are three $S$-domains per $k$-domain with moment directions rotated by 120$^{\circ}$ between $S$-domains \cite{magdom}. 

Note that the magnetic refinements reported below are based on the analysis of the magnetic Bragg reflections indexed by the four propagation vectors with half-integer indices mentioned above. Therefore, it concerns only the antiferromagnetic component of the magnetic arrangement. While it describes the whole magnetic structure in zero-field, it is accompanied by a rising ferromagnetic component when a magnetic field is applied, which is not refined in the present work.

\begin{figure}[h]
\begin{center}
\includegraphics[width=8cm]{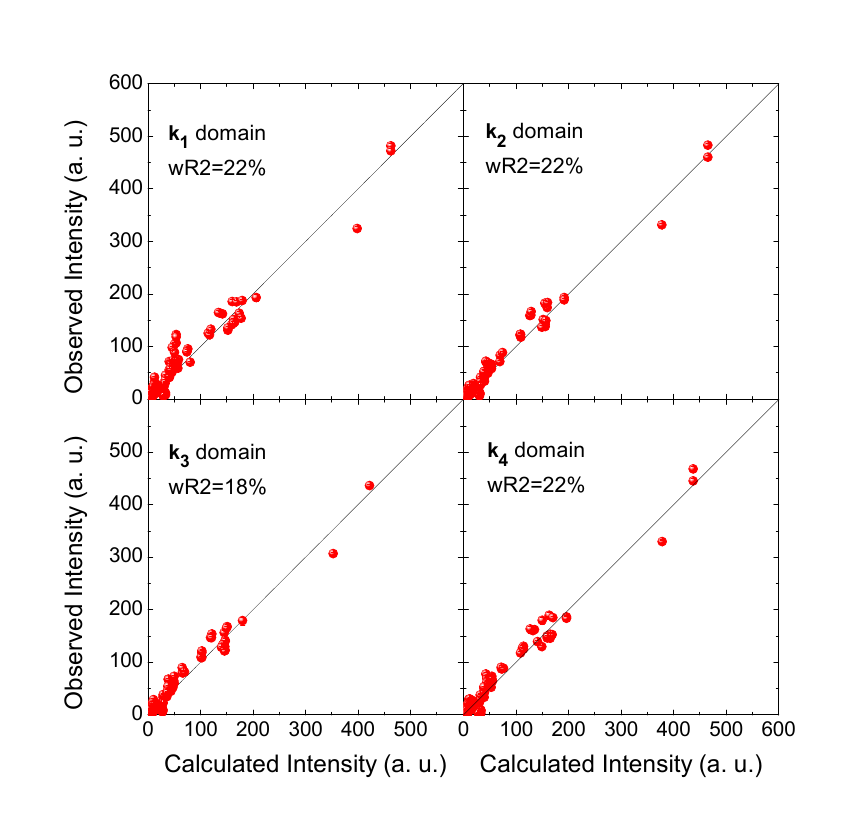}
\end{center}
\caption{Observed versus calculated intensity of the magnetic Bragg reflections measured for the four $k$-domains on D15 at $H$=0 T and $T$=10 K. The wR2 factors give the goodness of the fit.}
\label{SCRefinD15}
\end{figure}

\begin{figure}
\includegraphics[width=6cm]{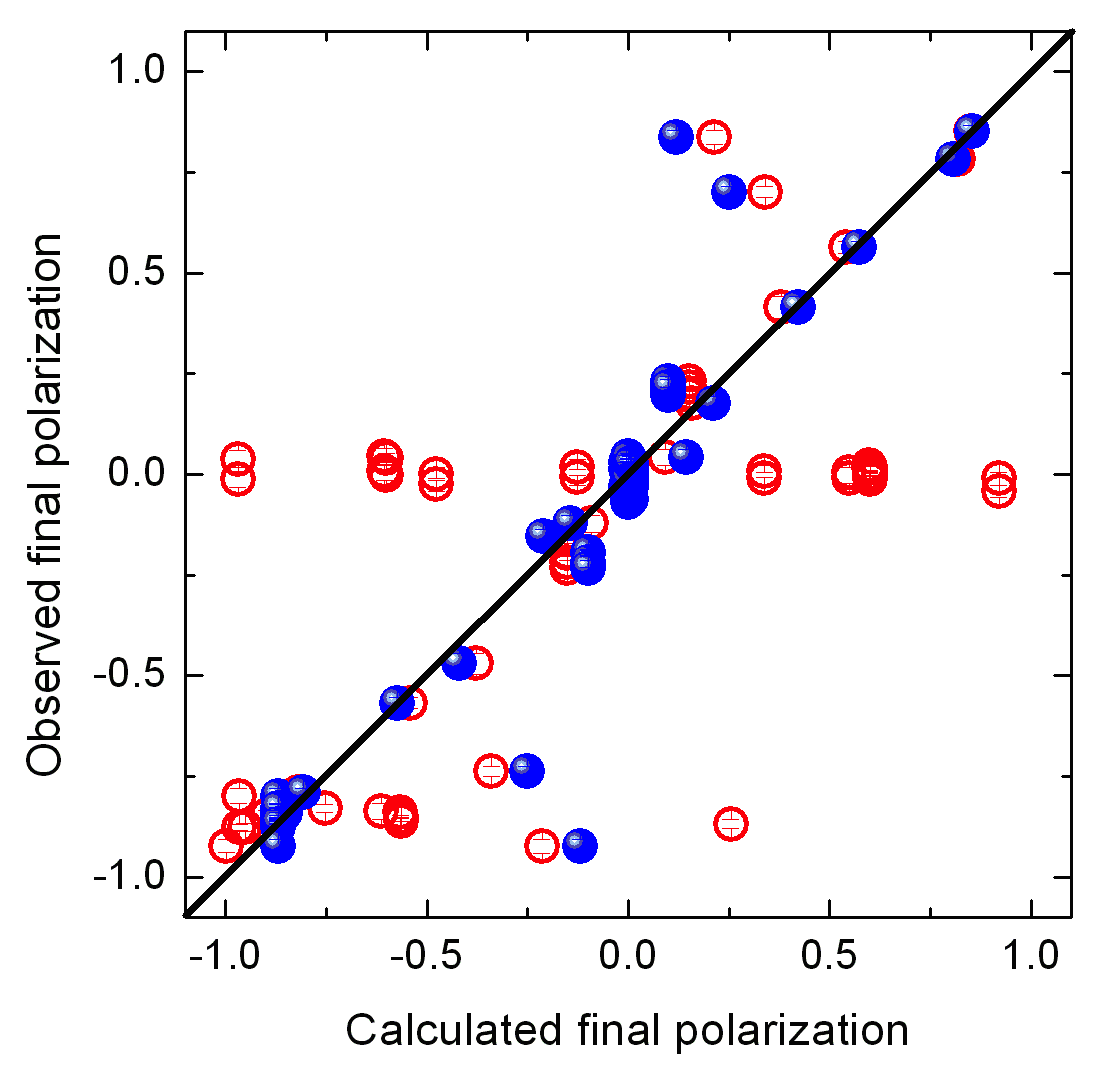}
\caption{Spherical polarization analysis at $T$=1.5 K using CRYOPAD on IN22: observed versus calculated components of the final polarization for 23 magnetic Bragg peaks belonging to the $\bf{k_1}$ domain. The magnetic moments on the TRI and KGM sites lie in the plane perpendicular to the [111] direction. The best fit is obtained when they are parallel to each other (blue filled circles, goodness of the fit wR2=21.4\%), and it deteriorates as soon as they depart from collinearity. For comparison, an example is given for perpendicular TRI and KGM spins (red empty circles, wR2=63.0 \%). Note that 4 points out of 207 are still not well described by the best model with collinear TRI and KGM magnetic moments. They corresponds to the P$_{YY}$ and P$_{ZZ}$ components of the polarization matrix for the (0.5, 0.5, 2.5) and (1.5, 1.5, -0.5) Bragg reflections. The origin of this discrepancy is unclear at the moment. }
\label{cryopad}
\end{figure}

In zero field, our refinement of the magnetic intensities collected on D15 at $T$=10 K (see figure \ref{SCRefinD15}) confirms the antiferromagnetic structure deduced from powder neutron diffraction \cite{Diaz2006,Matsuda2011}. It additionally proves that the TRI and KGM magnetic moments lie in the plane perpendicular to the $<$111$>$ directions given by the propagation vectors. Due to the presence of 12 domains that we found approximately equipopulated (see Table \ref{TableRefin}), the moment orientation could not be further determined using unpolarized neutrons. Polarized neutrons and spherical polarization analysis allowed us to determine the relative orientation of the magnetic moments on the TRI and KGM sites (but not their absolute direction that was only inferred from neutron scattering under magnetic field, as presented below). The best fit of the measured polarization matrix at 1.5 K for 23 reflections is shown with blue dots in figure \ref{cryopad}. It is obtained for a parallel orientation of TRI and KGM magnetic moments in the \{111\} planes (see figure \ref{Refin}a) and a population of the three $S$-domains equal to 26/40/34 \%. 

\begin{table}[h]
\caption{Results of the magnetic structure refinements in zero field (D15) and under applied magnetic fields (D23). In the successive columns are reported the value of the magnetic field in T, the $k$-domains population, the antiferromagnetic component of the magnetic moments in Bohr magneton on the KGM and TRI sites (only the amplitude at 0 T and the magnetic moment vector components under finite fields), and the population of the corresponding $S$-domains, listed as in figure \ref{MagDom}.}
\begin{center}
\begin{tabular}{|lc|ccccccc|}
\hline
     H        &  &       $k$-domain           &  &      M$_{KGM}$     &  &     M$_{TRI}$    &  &     $S$-domain     \\
\hline
 0   &  &	      $\bf{k_1}$ 0.255    &  &    3.3       &  &     2.2    &  &    0.34/0.28/0.38   \\
                                      &  &	      $\bf{k_2}$ 0.257    &  &    3.2       &  &     2.3    &  &    0.35/0.34/0.31   \\
                                      &  &	      $\bf{k_3}$ 0.234    &  &    3.2       &  &     2.2    &  &    0.35/0.34/0.31   \\
                                      &  &	      $\bf{k_4}$ 0.255    &  &    3.2       &  &     2.2    &  &    0.32/0.28/0.40   \\
 \hline
5.5   &  &	      $\bf{k_1}$ 0.510    &  &   (1.29, 1.29, -2.59)       &  &     0    &  &    0.97/0.015/0.015   \\
                                      &  &	      $\bf{k_2}$ 0.490    &  &    (1.27, 1.27, 2.53)       &  &     0    &  &    0.94/0.06/0.0   \\

\hline
 9.4   &  &	      $\bf{k_1}$ 0.510    &  &    (0.74, 0.74, -1.48)       &  &     0    &  &    0.99/0.0/0.01   \\
                                      &  &	      $\bf{k_2}$ 0.490    &  &    (0.71, 0.71, 1.43)       &  &     0    &  &    0.98/0.0/0.02   \\

\hline
12   &  &	      $\bf{k_3}$ 0.605    &  &    (1.10, 1.10, 0)       &  &     0    &  &    1/0/0   \\
                                      &  &	      $\bf{k_4}$ 0.395    &  &    (1.09, 1.09, 0)     &  &     0    &  &    1/0/0   \\
\hline
\end{tabular}
\end{center}
\label{TableRefin}
\end{table}


\begin{figure}[h]
\begin{center}
\includegraphics[width=8cm]{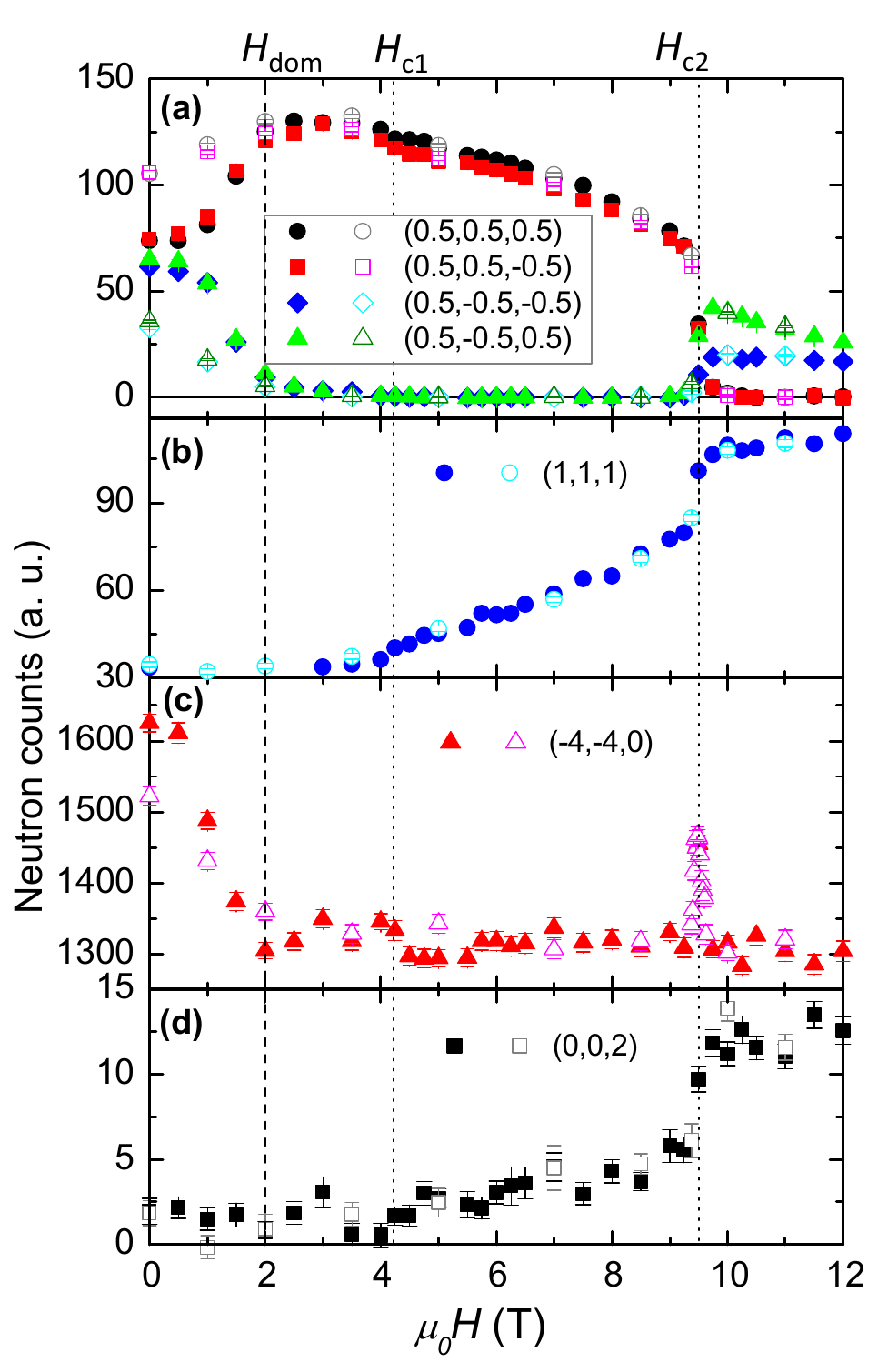}
\end{center}
\caption{Neutron counts at the peak maximum versus magnetic field applied along the [1-10] direction at 4.5 K for: (a) four magnetic Bragg reflections representative of the AFM $k$-domains, (b) the (1, 1, 1) weak nuclear reflection on top of which grows a field-induced FM component, (c) the (-4, -4, 0) strong nuclear reflection sensitive to the extinction, (d) the (0, 0, 2) forbidden nuclear reflection. The $\it{H_{\rm dom}}$, $\it{H_{\rm c1}}$ and $\it{H_{\rm c2}}$ characteristic fields are indicated by vertical lines. The filled (open) symbols correspond to increasing (decreasing) field measurements.}
\label{Magdom}
\end{figure}

\begin{figure}[h]
\begin{center}
\includegraphics[width=8cm]{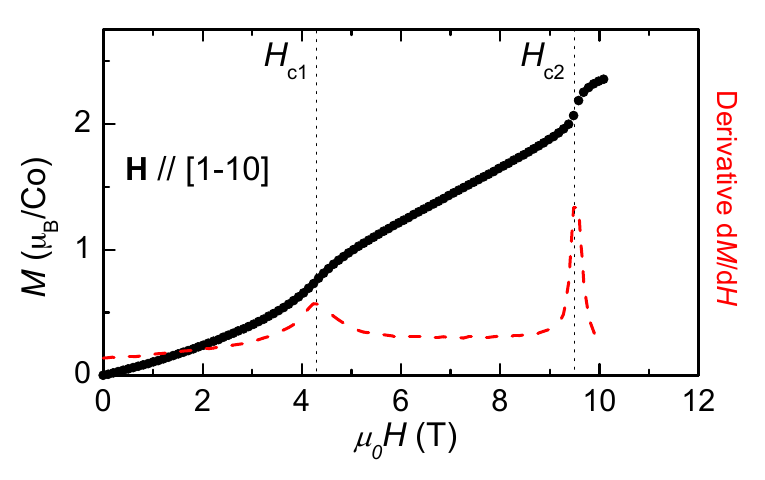}
\end{center}
\caption{Single-crystal magnetization $M$ and its derivative (red dashed line) measured as a function of the magnetic field oriented along the [1-10] direction at 2 K.}
\label{Mag}
\end{figure}

Under a magnetic field, the populations of the magnetic domains are expected to vary, the domains with the AFM component perpendicular to the magnetic field being favored by the exchange and Zeeman terms. We followed four AFM reflections while applying a magnetic field along the 2-fold axis [1-10] direction: (0.5, 0.5, 0.5), (0.5, 0.5, -0.5), (0.5, -0.5, -0.5), (0.5, -0.5, 0.5). They are characteristic of the four $k$-domain populations and have roughly the same intensity at $H$=0. As shown in figure \ref{Magdom}a, when increasing the magnetic field, the 4 equipopulated $k$-domains start to split into two pairs of reflections with different intensities, one pair being selected ($\bf{k_1}$, $\bf{k_2}$) whereas the other pair ($\bf{k_3}$, $\bf{k_4}$) is suppressed with increasing field. This selection occurs irreversibly mostly below $\mu_0H_{\rm dom}\approx$ 2 T. At $\mu_0H_{\rm c1}$=4.3 T, there is a slight change of curvature of the field dependence of the (0.5, 0.5, 0.5) and (0.5, 0.5, -0.5) signal. This field corresponds to the first anomaly observed in the magnetization curve measured on a single-crystal shown in figure \ref{Mag}, in agreement with the results of Hoshi {\it et al.} \cite{Hoshi2007}. The second magnetization anomaly occurs at $\mu_0H_{\rm c2}$=9.5 T, when the populations of the four $k$-domains change again abruptly in a remarkable way: the two selected domains ($\bf{k_1}$, $\bf{k_2}$) vanish and the two $k$-domains that were absent ($\bf{k_3}$, $\bf{k_4}$) are populated (see figure \ref{Magdom}a). 

\begin{figure}[h]
\begin{center}
\includegraphics[width=8cm]{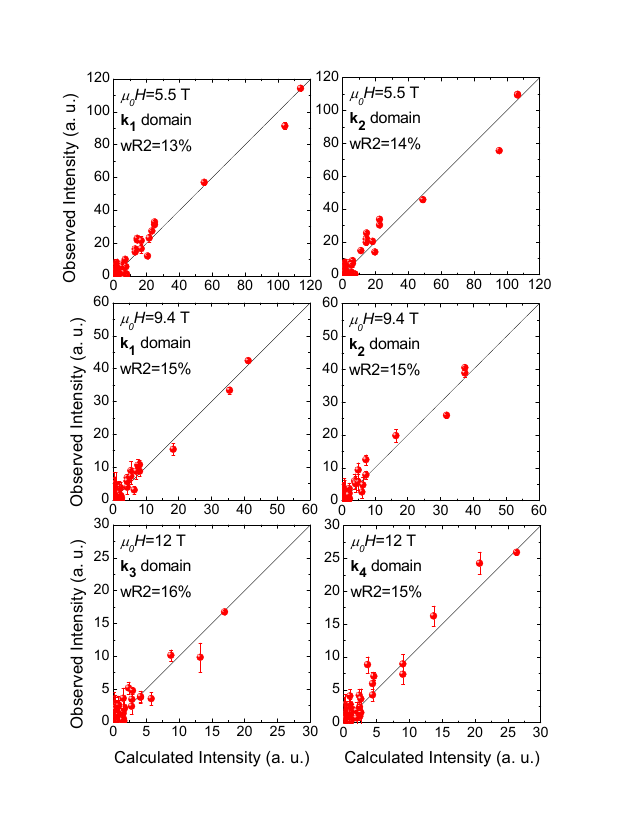}
\end{center}
\caption{Observed versus calculated intensity of the magnetic Bragg reflections measured on D23 at T=4.5 K, and refined at 5.5 T, at 9.4 T ({\it H$_{\rm c1}$$<$H$<$H$_{\rm c2}$}), and at 12 T ({\it H$>$H$_{\rm c2}$}). The wR2 factors give the goodness of the fit.}
\label{SCRefinD23}
\end{figure}

In addition to the AFM components probed with the previous reflections, a gradual rise of the intensity of the (1, 1, 1) nuclear reflection is observed (see figure \ref{Magdom}b). It is associated with the FM component ($M$) developing along the field on the TRI and KGM sites. It leads to the integrated magnetic intensity on the (1, 1, 1) reflection, given by the square of the magnetic structure factor, $ \propto (M_{TRI}-3M_{KGM})^2$ \cite{Matsuda2011}. Two step-like increases of this signal are observed, a small one at $\it{H_{\rm c1}}$ and a pronounced one at $\it{H_{\rm c2}}$. Above $\it{H_{\rm c2}}$, both a FM and an AFM contributions are thus present indicating a canted magnetic structure, consistent with the magnetization still increasing in higher fields \cite{Diaz2006}.


The field-induced antiferromagnetic component of the magnetic orders could be determined from refinements of the half integer magnetic reflections collected on D23 at $T$=4.5 K under the selected fields of 5.5, 9.4 and 12 T as shown in figure \ref{SCRefinD23}. At $\mu_0H$=5.5 and 9.4 T, i.e. for {\it H$_{\rm c1}$$<$H$<$H$_{\rm c2}$}, the KGM AFM component of the moments is found aligned along the [11-2] direction for the $\bf{k_1}$ selected domain (see figure \ref{Refin}b) and along the [112] direction for the $\bf{k_2}$ selected domain. The amplitude of this component decreases from 3.1 to 1.8 $\mu_B$ between 5.5 and 9.4 T while the FM component increases. On the TRI site, the AFM component is found equal to zero above $\it{H_{\rm c1}}$, as already proposed \cite{Matsuda2011}. At $\mu_0H$=12 T, i.e. for {\it H$>$H$_{\rm c2}$}, the KGM AFM component of the two new $\bf{k_3}$ and $\bf{k_4}$ selected domains is found aligned along a unique direction [110], also perpendicular to the applied magnetic field (see figure \ref{Refin}c). At this field, the AFM and FM components of the KGM magnetic moment are found $\approx$ 1.5 and 2.4 $\mu_B$ respectively from neutron diffraction refinement and magnetization measurements. 

The field-induced magnetic behavior is summarized in Table \ref{Tabledomain} and leads to the following scenario: at zero-field, we assume that both the TRI and KGM moments are parallel to each other and along the 12 equivalent $<$112$>$ directions. Increasing the field applied along [1-10] selects, above $\it{H_{\rm dom}}$, the two [11-2] and [112] directions perpendicular to it, and thus the two associated $\bf{k_1}$ and $\bf{k_2}$ domains. $\it{H_{\rm c1}}$ marks an AFM to FM transition of the TRI magnetic moments that become oriented along the field. At $\it{H_{\rm c2}}$, a new magnetic order is stabilized with the AFM component on the KGM sites aligned along the [110] direction. This orientation is compatible with the selection of two new domains, $\bf{k_3}$ and $\bf{k_4}$, and implies a redefinition of the KGM and TRI planes. A canted magnetic structure is induced by the field on the former while the latter are ferromagnetic.

\begin{table} [htb]
\centering
\caption{Selection of the antiferromagnetic domains under a magnetic field applied along [1-10]. For the three field ranges given in the Table, the columns indicate the selected propagation vectors and the orientation of the antiferromagnetic component of the KGM and TRI magnetic moments. Note that in zero field the four magnetic domains are equipopulated and that they get selected for $\it{H}>\it{H_{\rm dom}}$. Above $\it{H_{\rm c1}}$, the TRI moments are ferromagnetically aligned along the field. }
\begin{tabular}{ccc}
\hline
\hline
$\it{H_{\rm dom}<H<H_{c1}}$       &  $\it{H_{\rm c1}<H<H_{c2}}$     &  $\it{H_{\rm c2}<H}$ \\
\hline
    \begin{tabular}{ccc}  $k$   &   KGM      &     TRI  \\ \hline $\bf{k_1}$       &  [11-2]      &  [11-2]    \\  $\bf{k_2}$   &   [112]   &  [112] \end{tabular} 
& \begin{tabular}{ccc}  $k$   &   KGM      &     TRI  \\ \hline $\bf{k_1}$       &  [11-2]      &  none   \\   $\bf{k_2}$    &  [112]    &  none \end{tabular} 
& \begin{tabular}{ccc}  $k$   &   KGM      &     TRI   \\ \hline $\bf{k_3}$       &  [110]      &  none   \\   $\bf{k_4}$    &  [110]    &  none\end{tabular}  \\
\hline
\end{tabular}
\label{Tabledomain}
\end{table}

\subsection{Structural changes}

Correlated to the complex magnetic behavior of GeCo$_2$O$_4$, there is a remarkable variation of the extinction. It corresponds to a decrease of the intensity of strong Bragg reflections when the structural crystal quality is improved possibly due to magnetostructural effects \cite{Becker1974}. An irreversible decrease of the intensity is observed on the intense nuclear reflection (-4, -4, 0) up to $\it{H_{\rm dom}}$ (see figure \ref{Magdom}c), which shows that the structural quality is improved when the magnetic domains are selected. The intensity is then constant up to $\it{H_{\rm c2}}$ where a very narrow peak of extra intensity is visible, indicating a transient crystal deterioration at the transition toward the new magnetic structure. 

\begin{figure}[h]
\begin{center}
\includegraphics[width=8cm]{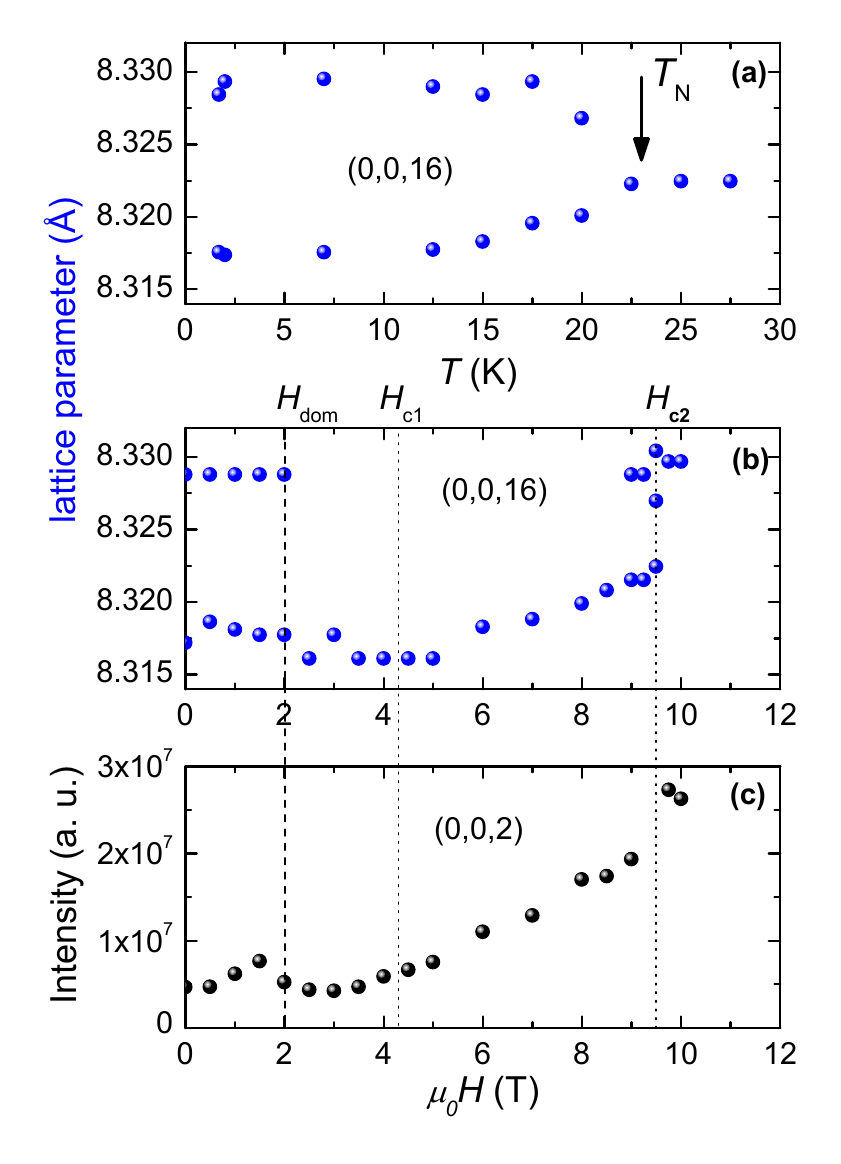}
\end{center}
\caption{Single-crystal X-ray diffraction: temperature dependence in zero field of the lattice parameter determined from the (0, 0, 16) reflection (a). Field dependence at 2 K under a magnetic field along the [1-10] direction of the lattice parameter determined from the (0, 0, 16) reflection (b), of the intensity of the (0, 0, 2) forbidden reflection (c). The $\it{H_{\rm dom}}$, $\it{H_{\rm c1}}$ and $\it{H_{\rm c2}}$ characteristic fields are indicated by vertical lines.}
\label{RX}
\end{figure}

\begin{figure}[h]
\begin{center}
\includegraphics[width=5cm]{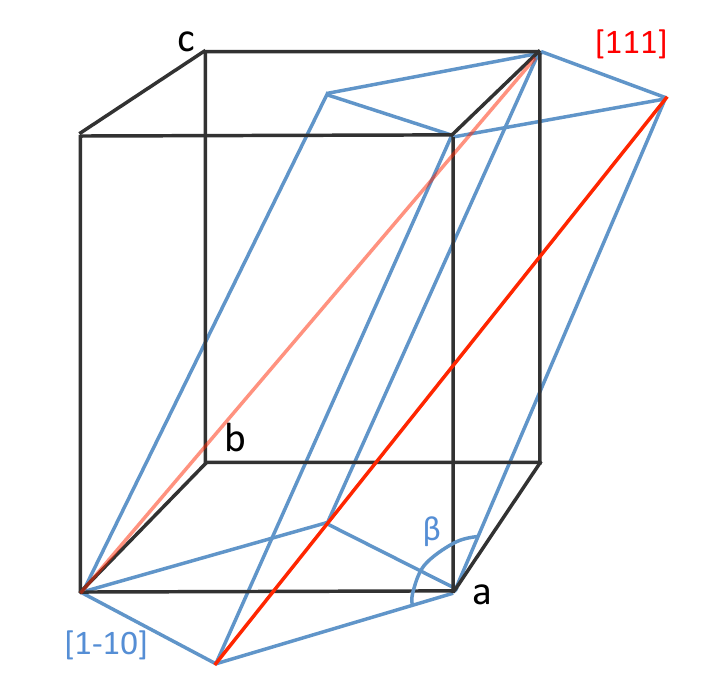}
\end{center}
\caption{Structural distortion: example of monoclinic cell (in blue), with a unique 2-fold axis [1-10], derived from the tetragonally distorted cubic cell (black) with the elongated $c$ axis. The monoclinic distortion is achieved through an additional trigonal distortion (here along the [111] direction) leading to a deviation of the $\beta$ angle from 125.264$^{\circ}$.}
\label{MagStruc}
\end{figure}

\begin{table} [htb]
\centering
\caption{Description of the magnetostructural domains. The successive columns indicate the tetragonal domains characterized by their elongated axis, the 12 monoclinic domains characterized by the trigonal axis of the cubic cell and the monoclinic 2-fold axis, and the associated 24 magnetic domains for the magnetic moments along the $<112>$ direction, with the additional requirement that the magnetization must be the closest to the plane perpendicular to the tetragonal axis. The domains selected by a magnetic field applied along the [1-10] direction are in bold. }
\begin{tabular}{ccc}
\hline
\hline
Tetragonal       &  Monoclinic    &  Magnetic \\ \hline
      $a$   &  \begin{tabular}{cc}  [111]     &   [01-1]  \end{tabular}      &    \begin{tabular}{cc}  {\bf [11-2]}   &   [1-21]   \end{tabular} \\
               &  \begin{tabular}{cc}  [11-1]   &   [011]    \end{tabular}      &    \begin{tabular}{cc}  {\bf [112]}    &   [1-2-1]  \end{tabular} \\ 
               &  \begin{tabular}{cc}  [1-1-1]  &   [01-1]  \end{tabular}      &    \begin{tabular}{cc}  [1-12]   &   [12-1]  \end{tabular} \\ 
               &  \begin{tabular}{cc}  [1-11]   &   [011]    \end{tabular}      &    \begin{tabular}{cc}  [1-1-2]  &  [121]     \end{tabular} \\ \hline
     $b$    &  \begin{tabular}{cc}  [111]     &   [10-1]  \end{tabular}      &    \begin{tabular}{cc}  [2-1-1]  &    {\bf [11-2]}  \end{tabular} \\
               &  \begin{tabular}{cc}  [11-1]   &   [101]    \end{tabular}      &    \begin{tabular}{cc}  [2-11]   &    {\bf [112]}   \end{tabular} \\ 
               &  \begin{tabular}{cc}  [1-1-1]  &   [101]  \end{tabular}        &    \begin{tabular}{cc}  [211]    &   [1-12]  \end{tabular} \\
               &  \begin{tabular}{cc}  [1-11]   &   [10-1]    \end{tabular}     &    \begin{tabular}{cc}  [21-1]   &   [1-1-2]   \end{tabular} \\ \hline
     $c$    &  \begin{tabular}{cc}  [111]     &   [1-10]  \end{tabular}       &    \begin{tabular}{cc}  [2-1-1]  &   [1-21]  \end{tabular} \\ 
               &  \begin{tabular}{cc}  [11-1]   &   [1-10]   \end{tabular}      &    \begin{tabular}{cc}  [2-11]    &   [1-2-1]   \end{tabular} \\ 
              &  \begin{tabular}{cc}   [1-1-1]  &   [110]  \end{tabular}        &    \begin{tabular}{cc}  [211]     &   [12-1]  \end{tabular} \\ 
              &  \begin{tabular}{cc}   [1-11]   &   [110]    \end{tabular}      &    \begin{tabular}{cc}   [21-1]   &   [121]   \end{tabular} \\ \hline
\end{tabular}
\label{TableMagStruc}
\end{table}

In the present paper, these magnetostructural effects were further investigated by single-crystal X-ray diffraction on ID06. Back to the temperature dependence in zero field, the (0, 0, 16) structural reflection splits into two peaks below $T_{\rm N}$ (see figure \ref{RX}a). This agrees with a main cubic-to-tetragonal structural distortion implying a 0.12 \% elongation of one of the cubic axes \cite{Hoshi2007}, hence yielding three tetragonal domains (see Table \ref{TableMagStruc} and the Supplemental Material for complementary powder neutron diffraction results). 

It should be noted however that, in a single-$k$ picture, a (1/2, 1/2, 1/2) propagation vector for the magnetic structure cannot result in a tetragonal strain. An important issue is therefore whether the magnetic order is described by only one or several members of the star of the $k$-vector. A similar puzzling behavior was reported in CoO: this oxide was known to undergo a cubic-to-tetragonal structural transition at the temperature of the collinear magnetic ordering, itself described by an AFM stacking of ferromagnetic planes with a (1/2, 1/2, 1/2) propagation vector and the magnetic moment along the $<$112$>$ directions \cite{Ressouche2006}. Symmetry arguments \cite{Herrmann1978} were invoked showing that, if the tetragonal distortion is driven by the magnetic ordering, the magnetic structure has to be described by 4 propagation vectors, leading to a non-collinear magnetic arrangement. On the other hand, if the main tetragonal distortion is not caused by the magnetic structure, a single-$k$ magnetic structure will induce an additional small trigonal distortion leading to a monoclinic space group. In CoO, the evolution of the magnetic domains under applied stress allowed to discriminate between the two scenarios, in favor of a single-$k$ structure \cite{Herrmann1978}. Direct evidence of a small but finite trigonal lattice distortion was finally found through X-ray diffraction \cite{Jauch2001}, which agreed with the magnetic structure symmetry. 

In GeCo$_2$O$_4$, as in CoO, a tetragonal distortion resulting from the magnetic ordering would imply 4 propagation vectors and a non-collinear magnetic structure. This is hardly compatible with the field-induced magnetic domain selection that we observe and that we can explain straightforwardly in a single-$k$ magnetic structure. Similarly to CoO \cite{Herrmann1978,Jauch2001}, we therefore suggest that the single-$k$ magnetic structure emerges in a primarily tetragonal context, inducing an additional small trigonal distortion (see figure \ref{MagStruc}). This should lead to a monoclinic space group, that has still to be evidenced \cite{footnote2}. This scenario is consistent with the reported first-order transition at $T_{\rm N}$ in GeCo$_2$O$_4$ \cite{Lashley2008,Hubsch1987}. Also supporting this assumption is the orientation of the Co$^{2+}$ magnetic moments that we determined along the same direction as in CoO \cite{Ressouche2006}. The monoclinic distortion is characterized by a unique 2-fold axis along the $<110>$ direction perpendicular to the selected cubic trigonal $<111>$ direction and to the tetragonal axis. In this case, each tetragonal domain is further split into four monoclinic domains, yielding 12 structural domains (see Table \ref{TableMagStruc}).

It is interesting at this stage to correlate the structural and the magnetic domains. For this purpose, we have studied the field evolution of the (0, 0, 16) nuclear Bragg peak that got split below $T_{\rm N}$ in zero field. Under a magnetic field along [1-10], the two (0, 0, 16) peaks are still visible up to $\mu_0\it{H_{\rm dom}}$=2 T, but only one peak remains above $\it{H_{\rm dom}}$ associated with the smaller lattice parameter (see figure \ref{RX}b). It means that the field selects two tetragonal domains out of three, with an $a$ or $b$ elongated axis. To understand the relation between this field-induced structural domain selection and the magnetic domain selection, we refer to the behavior of the magnetic moment of Co$^{2+}$ in CoO films submitted to different strains that compress or elongate the tetragonal axis. It has been shown in particular that for an elongated tetragonal axis, the magnetic moment should lie in the plane perpendicular to this axis \cite{Csiszar2005}. This condition in GeCo$_2$O$_4$ allows to identify 24 magnetostructural domains listed in Table \ref{TableMagStruc}. The trigonal cubic axis characterizing the distortion coincides with the direction of the magnetic propagation vector. Our single-crystal neutron diffraction results indicate that $\bf{H}\parallel$~[1-10] selects, below $\it{H_{\rm c2}}$, the magnetostructural domains with the [112] and [11-2] magnetic moments directions. These are indeed closer to the (100) and (010) planes than to the (001) one, hence compatible with the $a$ and $b$ elongated axis, as observed. 

From $\it{H_{\rm c1}}$ to $\it{H_{\rm c2}}$, the position of the remaining peak progressively changes signaling an increase in the $c$ lattice parameter. Above $\it{H_{\rm c2}}$, a unique peak finally appears abruptly shifted at a value corresponding to a larger $c$ lattice parameter. In addition to the lattice parameter variation, other structural changes start to occur for $\it{H>H_{c1}}$. For instance, the (0, 0, 2) structural reflection, forbidden by the $4_1/d$ symmetry element of the Fd-3m space group, rises in intensity from $\it{H_{\rm c1}}$ to $\it{H_{\rm c2}}$ where it presents a marked step, as observed both with neutron and X-ray diffraction (see figures \ref{Magdom}d and \ref{RX}c) \cite{footnote3}. 

\section{Discussion}

These results provide a strong indication that the field-induced evolution of the magnetic structure of GeCo$_2$O$_4$ is not merely due to an energy balance between the Zeeman term, the magnetic exchange interactions and the single ion magnetocristalline anisotropy. The key ingredient is actually the magnetostructural coupling. The TRI and KGM sites, which were coupled in zero-field through a structural distortion at $T_{\rm N}$, get decoupled above $\it{H_{\rm c1}}$ when the field strength is enough to polarize the TRI planes. This renders ineffective the zero-field structural distortion in reducing the magnetic frustration and lowering the energy of the system. It thus triggers new structural changes evidenced by the rise of the (0, 0, 2) structural reflection and the increase of the $c$ lattice parameter. These magnetoelastic effects ultimately lead at $\it{H_{\rm c2}}$ to an abrupt structural change and to a novel magnetic configuration. The latter consists in a stacking of FM planes of KGM moments. The canted plane to plane arrangement exhibits a ferromagnetic and an antiferromagnetic components, both perpendicular to the $<$111$>$ directions. The AFM component is now orientated along the [110] direction, which implies the magnetic domain switching from $\bf{k_1}$ and $\bf{k_2}$ to $\bf{k_3}$ and $\bf{k_4}$ at $\it{H_{\rm c2}}$, and a change in the KGM and TRI sites distribution. 

This remarkable spin reorientation can be ascribed to a change of the magnetocrystalline anisotropy of the Co$^{2+}$ ions induced by the structural deformation acting on the Crystal Field parameters. The exact nature of the structural changes occurring at $\it{H_{\rm c2}}$ cannot be unambiguously established. From an analogy with the Cr spinels, the first possibility is that this is a transition towards a new structure of cubic symmetry with a lattice parameter enlarged with respect to the paramagnetic state. The second option rather implies a change of the selected structural domains, triggered by the magnetism. The weakening of the structural distortion tilts the magnetic moments from the [112] and [11-2] directions to the (001) plane. The magnetic moments are then perpendicular to the $c$-axis thus favoring, above $\it{H_{\rm c2}}$, the tetragonal structural domain with the elongated $c$-axis instead of those with elongated $a$ and $b$ axes selected for $\it{H_{\rm dom}<H<H_{\rm c2}}$. This is supported by the fact that the $c$ lattice parameter above $\it{H_{\rm c2}}$ is identical to the one of the zero-field domain that disappears above $\it{H_{\rm dom}}$. Finally, complementary studies are crucially needed, first to confirm the single-$k$ nature of the magnetic structure and the monoclinic distortion, and then to decide between the two above scenarios.

\section{Conclusion}

In conclusion, the spinel compound GeCo$_2$O$_4$ exhibits common features to both Cr/V spinels and binary oxides such as CoO, in spite of different interaction schemes or different topology of the lattice. Our neutron and X-ray single-crystal diffraction study has shown that unconventional behaviors are generated by the strong interplay between the structural and frustrated magnetic degrees of freedom under the influence of an external magnetic field. The high field transition is particularly rich, involving deep structural changes and a reshuffling of the inhomogeneous magnetization distribution, associated with a rare field-induced change of magnetic anisotropy. 

We thank J. Debray for the single-crystal orientation and polishing necessary for the X-ray experiment, E. Lhotel, R. Ballou and J. Robert for fruitful discussions, and L. Chaix, S. Diaz and A. deMuer for their help during the single-crystal neutron diffraction experiments.

\widetext
\clearpage
\begin{center}
\textbf{\large Supplemental material for "Field driven magnetostructural transitions in GeCo$_2$O$_4$"}
\end{center}

\onecolumngrid

\section{Magnetization versus temperature}

\begin{figure}[h]
\begin{center}
\includegraphics[width=9cm]{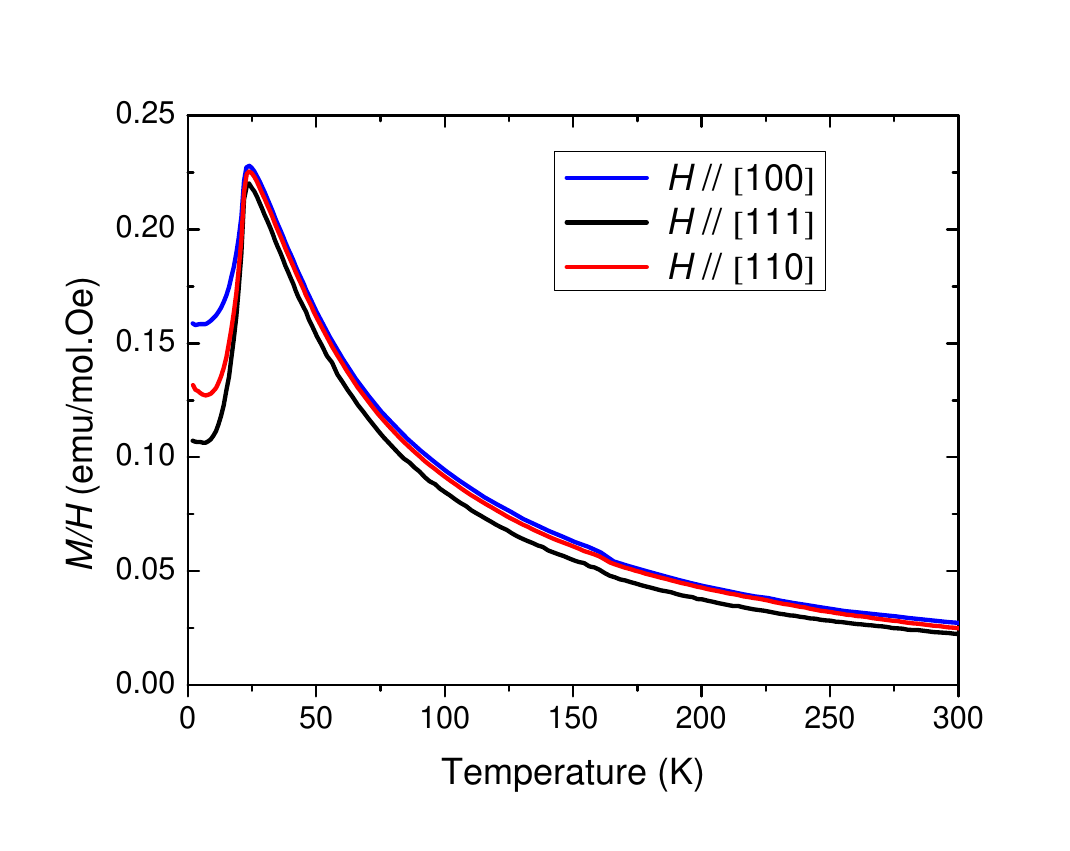}
\end{center}
\caption{Temperature dependence of the magnetic susceptibility (magnetization divided by the magnetic field, $M/H$, in the linear regime) measured on the GeCo$_2$O$_4$ single-crystal used in the neutron diffraction experiments, for three orientations of the magnetic field $\mu_0 H$=1 T with respect to the main crystallographic directions. The cusp in the three curves indicates the N\'eel temperature, $T_{\rm N}$=23.5 K, at which the antiferromagnetic ordering occurs.}
\label{D2B}
\end{figure}

\section{Powder neutron diffraction}

In addition to single-crystal X-ray diffraction, we performed powder neutron diffraction under magnetic field to investigate the structural distortion. Powder patterns were recorded on the D2B high-flux diffractometer of the Institut Laue Langevin (ILL) under a magnetic field up to 6 T on a pressed pellet of GeCo$_2$O$_4$ at the wavelength $\lambda$=1.594 \AA. Our powder neutron measurements are in agreement with the proposed cubic-to-tetragonal main distortion with an elongated tetragonal axis occurring at $T_{\rm N}$ \cite{Hoshi2007}: The large $2\theta$ angle  (0, 0, 8) nuclear reflection splits into 2 below $T_{\rm N}$ with the lower angle reflection being twice smaller in intensity than the higher angle reflection (see figure \ref{D2B}). Under a 6 T magnetic field, only the higher $2\theta$ angle peak is still visible which corresponds to the selection of tetragonal domains with an elongated $a$ or $b$ axis. 

\begin{figure}[h]
\begin{center}
\includegraphics[width=9cm]{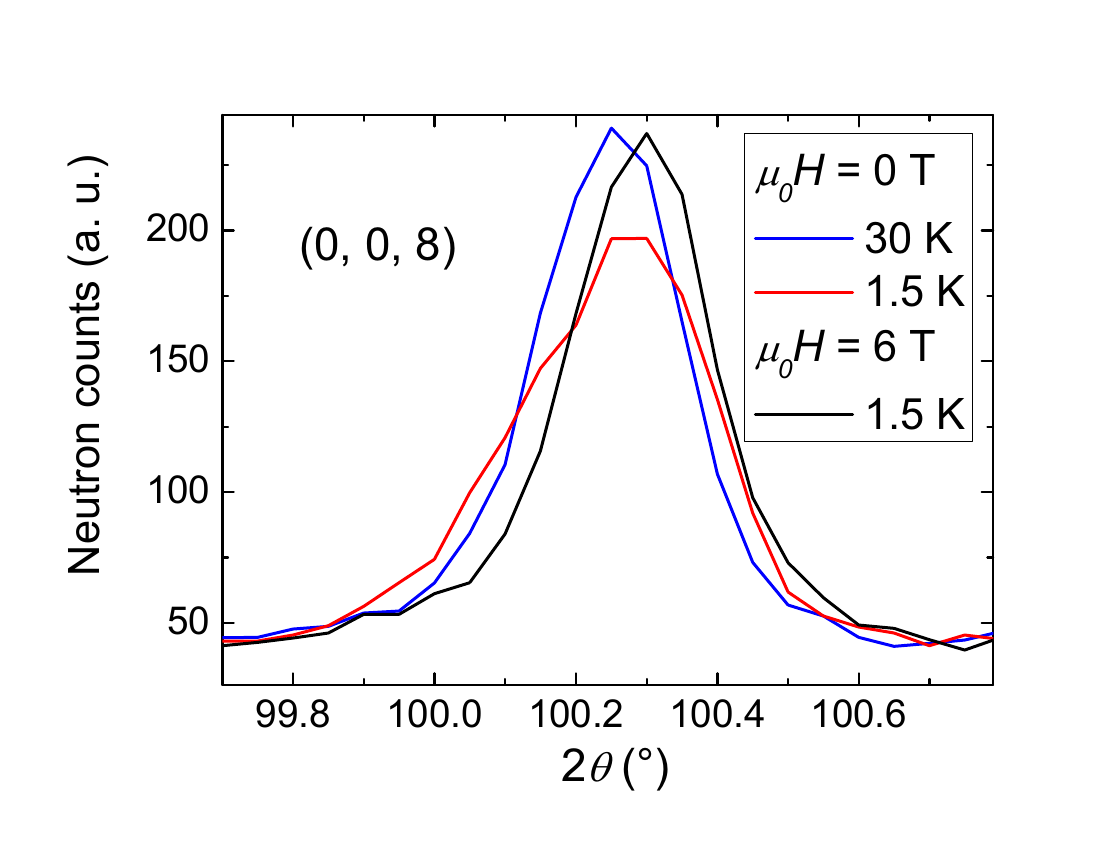}
\end{center}
\caption{Splitting of the (0, 0, 8) nuclear reflection at $T_{\rm N}$ from powder neutron diffraction measurements on D2B.}
\label{D2B}
\end{figure}



\begin{references}
\bibitem{magdom} 3 $S$-domains per $k$-domains are expected for parallel KGM and TRI magnetic moments, which is validated by the refinement. Our measurements are not sensitive to the 180$^{\circ}$ domains, which are thus ignored.
\bibitem{footnote2}The Bragg reflections of the type (0, 0, $\ell$), measured during the X-ray diffraction experiment, do not allow to estimate any additional trigonal distortion superimposing the main tetragonal one.
\bibitem{footnote3} The (0, 0, 2) reflection measured in neutron diffraction has both a nuclear and a FM contribution.
\end{references}
\end{document}